\documentclass{article}      
\usepackage{epsfig}
\title{ Solitons of the Resonant Nonlinear
Schrodinger Equation with Nontrivial Boundary Conditions and Hirota Bilinear Method}  
\author{Jyh-Hao Lee$^1$ and Oktay K. Pashaev$^2$ \\$^1$ Institute of Mathematics, Academia Sinica\\
Taipei, 11529, Taiwan\\
$^2$ Department of Mathematics,
Izmir Institute of Technology \\
Urla-Izmir, 35430, Turkey}     

\begin{document}             

\maketitle                   

\begin{abstract}
 Physically relevant soliton
solutions of the resonant nonlinear Schrodinger (RNLS) equation
with nontrivial boundary conditions, recently proposed for
description of uniaxial waves in a cold collisionless plasma, are
considered in the Hirota bilinear approach. By the Madelung
representation, the model transformed to the reaction-diffusion
analog of the NLS equation for which the bilinear representation,
soliton solutions and their mutual interactions are studied.

\end{abstract}
\section{Introduction}

 Recently for description of low dimensional gravity (the Jackiw-Teitelboim model),
  and a medium response to the action of a
quasimonochromatic wave with complex amplitude $\psi(x,t)$, which
is slowly varying function of the coordinate and the time,  a
novel integrable version of NLS equation namely

\begin{equation}
i \frac{\partial \psi}{\partial t} + \frac{\partial^2
\psi}{\partial x^2} + {\Lambda \over 4} |\psi|^2 \psi = s
\frac{1}{|\psi|}\frac{\partial^2 |\psi|}{\partial x^2} \psi
\label{RNLS}
\end{equation}
was introduced \cite{PashaevLeeMPLA}. This has been termed the
{\it resonant nonlinear Schrodinger} (RNLS) equation. It can be
considered as a third version of the NLS equation, intermediate
between the defocusing and focusing cases. Even though the RNLS
model is integrable for arbitrary values of the coefficient $s$,
the critical value $s = 1$ separates two distinct regions of
behavior. Thus, for $s < 1$ the model is reducible to the
conventional NLS equation. However, for $s> 1$ it is not reducible
to the usual NLS equation, but rather to a reaction-diffusion (RD)
system. In this case the model exhibits resonance solitonic
phenomena \cite{PashaevLeeMPLA}.

The RNLS equation can be interpreted as a particular realization
of the NLS soliton propagating in the  so called ''quantum
potential" $U_{Q}(x) =  |\psi|_{xx}/ |\psi|$. This potential,
responsible for producing  the quantum behavior, was introduced by
de Broglie \cite{deBroglie} and was subsequently used by Bohm
\cite{Bohm} to develop a hidden-variable theory in quantum
mechanics. It also appears in stochastic mechanics \cite{Nelson}.
Connections between such non-classical motions with the {\it
internal } spin motion and the {\it zitterbewegung} have been
considered in a series of papers (see \cite{Salesi}). Quantum
potentials also appear in proposed nonlinear extensions of quantum
mechanics with regard both to stochastic quantization
\cite{Guerra}, \cite{Smolin} and to corrections from quantum
gravity \cite{Bertolami}. It is noted that the RNLS equation, like
the conventional NLS equation, may also be derived in the context
of capillarity models \cite{Antonovskii}, \cite{Rogers}

Very recently \cite{LPRS} it was shown that the RNLS equation
appears in plasma physics, where it describes the propagation of
one-dimensional long magnetoacoustic waves in a cold collisionless
plasma subject to a transverse magnetic field. The complex wave
function satisfying RNLS equation is a combination of plasma
density and the velocity fields as in the Madelung representation.
The Backlund-Darboux transformations along with novel associated
nonlinear superposition principle were presented and used to
generate solutions descriptive of the interaction of solitonic
magnetoacoustic waves. This application requires to consider a
solution of the RNLS equation  with nontrivial boundary conditions
at infinity. The goal of the present paper is to derive such
solutions in the Hirota bilinear approach and study their mutual
interactions.

\section{Magnetoacoustic waves in cold plasma}

The dynamics of two-component cold collisionless plasma in the
presence of an external magnetic field $\bf B$
\cite{Karpman},\cite{Akhiezer}
 for uni-axial plasma propagation
\begin{equation}
{\bf u}  = u(x,t)\,{\bf e}_x,\,\,\,\,{\bf B}  = B(x,t)\,{\bf e}_z
\label{12}
\end{equation}
reduces to the form \cite{Adlam}

\begin{equation}
\frac{\partial \rho}{\partial t} + \frac{\partial}{\partial
x}(\rho u) = 0 , \label{contin2}
\end{equation}

\begin{equation}
\frac{\partial u}{\partial t} + u \frac{\partial u}{\partial x} +
\frac{B}{\rho}\frac{\partial B}{\partial x}  = 0 , \label{euler2}
\end{equation}
\begin{equation}
\frac{\partial }{\partial x}\left(\frac{1}{\rho}\frac{\partial
B}{\partial x}\right)  = B - \rho . \label{mag2}
\end{equation}
(where we set $B = 1$ and plasma density $\rho = 1$ at infinity)
This system is equivalent to that of Whitham and has also been
derived by Gurevich and Meshcherkin \cite{GurevichMesh}. It
describes the propagation of  nonlinear magnetoacoustic waves in a
cold plasma with a transverse magnetic field. It has been shown
recently by El, Khodorovskii and  Tyurina \cite{ElKhodorTyurina}
that
a system of the type  (\ref{contin2})-(\ref{mag2}) also occurs in
the context of hypersonic flow past slender bodies.

\section{A shallow water approximation}

Here,  we consider a  shallow water approximation to  the
magnetoacoustic system (\ref{contin2})-(\ref{mag2}). Thus,
rescaling the space and time variables via $x' = \beta x $ and $t'
= \beta t$, we have
\begin{equation}
\frac{\partial \rho}{\partial t'} + \frac{\partial}{\partial
x'}(\rho u) = 0 \label{contin3}
\end{equation}
\begin{equation}
\frac{\partial u}{\partial t'} + u \frac{\partial u}{\partial x'}
+ \frac{B}{\rho}\frac{\partial B}{\partial x'}  = 0\label{euler3}
\end{equation}
\begin{equation}
\beta^2\frac{\partial }{\partial
x'}\left(\frac{1}{\rho}\frac{\partial B}{\partial x'}\right)  = B
- \rho\label{mag3}
\end{equation}
On expansion of $B$ as a  power series in the parameter $\beta^2$
according to
\begin{equation}
B = \rho + \beta^2 b_2 (\rho, \rho_{x'}, \rho_{x'x'},...) +
O(\beta^4), \label{expand}
\end{equation}
insertion into  (\ref{mag3}) yields
\begin{equation}
b_2 = \frac{\partial }{\partial
x'}\left(\frac{1}{\rho}\frac{\partial \rho}{\partial x'}\right).
\label{btwo}
\end{equation}
Substitution of (\ref{expand}) into (\ref{euler3}) yields
\begin{equation}
\frac{\partial u}{\partial t'} + u \frac{\partial u}{\partial x'}
+ \frac{\partial \rho}{\partial x'} + \beta^2
\left[\frac{1}{\rho}\,\frac{\partial^3\rho}{\partial x'^3} -
\frac{2}{\rho^2} \frac{\partial \rho}{\partial x'}
\,\frac{\partial^2 \rho}{\partial x'^2} +
\left(\frac{1}{\rho}\,\frac{\partial \rho}{\partial x'}\right)^3
\right] = 0   \label{euler4}
\end{equation}
to  $O(\beta^2)$. Accordingly, the following system results:
\begin{equation}
\frac{\partial \rho}{\partial t'} + \frac{\partial}{\partial
x'}(\rho u) = 0 , \label{cold1}
\end{equation}
\begin{equation}
\frac{\partial u}{\partial t'} + u \frac{\partial u}{\partial x'}
+ \frac{\partial \rho}{\partial x'} + \beta^2
\frac{\partial}{\partial x'}
\left[\frac{1}{\rho}\,\frac{\partial^2 \rho}{\partial x'^2} -
\frac{1}{2}\left(\frac{1}{\rho}\,\frac{\partial \rho}{\partial
x'}\right)^2 \right] = 0 . \label{cold2}
\end{equation}
This describes the propagation of long magnetoacoustic waves in a
cold plasma of density $\rho$ moving with velocity $u$ across the
magnetic field as given by (\ref{12}), (\ref{expand})(see
\cite{GurevichKrylov87}, \cite{GurevichKrylov88}). In this system
the dispersion is negative, i.e., the wave velocity decreases with
increasing wave vector $k$.

\section{Resonant NLS}

Introducing the velocity potential $ S(x,t) = -\frac{1}{2}\int^{x}
u(x,t) dx $ so that $u = -2  \partial S/\partial x$ and
integrating once equation (\ref{cold2}) we get the system
\begin{equation}
\frac{\partial \rho}{\partial t} - 2\frac{\partial}{\partial
x}(\rho \frac{\partial S}{\partial x}) = 0 \label{madelung1}
\end{equation}
\begin{equation}
-\frac{\partial S}{\partial t} +  \left(\frac{\partial S}{\partial
x}\right)^2 + \frac{1}{2}\rho + \frac{\beta^2}{2}
\left[\frac{1}{\rho}\,\frac{\partial^2 \rho}{\partial x^2} -
\frac{1}{2}\left(\frac{1}{\rho}\,\frac{\partial \rho}{\partial
x}\right)^2 \right] = 0\label{madelung2}
\end{equation}
(here we skip prime superscript). Combining $\rho$ and $S$ as one
complex function
\begin{equation}
\psi = \sqrt{\rho} e^{-i S} \label{compfunc}
\end{equation}
the system (\ref{madelung1}),(\ref{madelung2}) can be represented
as the nonlinear Schrodinger equation with quantum potential, the
RNLS (\ref{RNLS}),
\begin{equation}
i \frac{\partial \psi}{\partial t} + \frac{\partial^2
\psi}{\partial x^2} - \frac{1}{2} |\psi|^2 \psi = (1 + \beta^2)
\frac{1}{|\psi|}\frac{\partial^2 |\psi|}{\partial x^2} \psi
\label{RNLSdefocusing}
\end{equation}
where $\Lambda = -2$, $s = 1 + \beta^2$, and  we have used
expression for the quantum potential \begin{equation}
\frac{1}{\rho}\,\frac{\partial^2 \rho}{\partial x^2} -
\frac{1}{2}\left(\frac{1}{\rho}\,\frac{\partial \rho}{\partial
x}\right)^2 = 2\frac{1}{\sqrt{\rho}}\frac{\partial^2
\sqrt{\rho}}{\partial x^2}\label{quantumpotential}
\end{equation}
 Since parameter $s
> 1$ equation (\ref{RNLSdefocusing}) can not be transformed to the  NLS \cite{PashaevLeeMPLA}. But by
combining $\rho$ and $S$ as a couple of real functions
\begin{equation}
e^{(+)} = \sqrt{\rho}\, e^{\frac{1}{\beta}S},\,\,\,e^{(-)} =
-\sqrt{\rho}\, e^{-\frac{1}{\beta}S}
 \end{equation}
 so that $e^{(+)} > 0$, $e^{(-)} < 0$,
equations (\ref{madelung1}),(\ref{madelung2})
 can be written as the reaction-diffusion (RD) system
\begin{equation}
\mp \frac{\partial e^{(\pm)}}{\partial \tau} + \frac{\partial^2
e^{(\pm)}}{\partial x'^2} - \frac{1}{2\beta^2} e^{(+)} e^{(-)}
e^{(\pm)} = 0 \label{RD}
 \end{equation}
 where $\tau = \beta t'$.

 Linearization of (\ref{RNLSdefocusing}) near the "condensate"
 solution $\psi = \sqrt{\rho_0} e^{-i\rho_0 t/2}$ gives dispersion
 $
 \omega = \sqrt{\rho_0}\, k \,\sqrt{1 - \frac{\beta^2}{\rho_0} \,
 k^2}$.
This dispersion is negative, i.e. the wave velocity decreases with
increasing wave vector $k$ and is unstable for short waves with $k
> k_{cr}$, where $k_{cr} = \sqrt{\rho_0}/\beta$. However this
instability results from the truncation of the dispersion relation
\begin{equation}
 \omega^2 = \rho_0 \frac{ k^2}{1 + \frac{\beta^2}{\rho_0} \,
 k^2}\label{dispersion}
 \end{equation}
for the system (\ref{contin3}), (\ref{euler3}), (\ref{mag3}) and
does not correspond to any actual physical effect for shallow
water waves \cite{ElGrimPavlov}. Though this system is linearly
stable for all wave numbers $k$, it is not known to be integrable.
Therefore, it is not as suitable for studying wave interactions as
the RNLS, which is completely integrable and admits a rich variety
of the exact solutions.

\section{Steady State Flow and Solitons}

Now we consider the system (\ref{cold1}),(\ref{cold2}) for the
steady state flow. It describes the motion with fixed velocity
$u(t,x) = u_0 = const$ for which the continuity equation
(\ref{cold1}) implies $ \frac{\partial \rho}{\partial t} +
u_0\frac{\partial  \rho }{\partial x} = 0$ or that fluid density
has the travelling wave form $\rho = \rho (x - u_0 t)$, where $x'
= \beta x $, $t' = \beta t$. Equation (\ref{cold2}) in this case
gives
\begin{equation}
 \rho  +
\left[\frac{1}{\rho}\,\frac{\partial^2 \rho}{\partial x^2} -
\frac{1}{2}\left(\frac{1}{\rho}\,\frac{\partial \rho}{\partial
x}\right)^2 \right] = a = const.\label{compensation}
\end{equation}
This has a simple physical interpretation. The motion with fixed
velocity implies that sum of all forces acting on the system is
zero. In our case Eq.(\ref{compensation}) describes compensation
of the nonlinearity  by the quantum potential. Using
(\ref{quantumpotential}) this equation gives
\begin{equation}
 \rho  + 2\frac{1}{\sqrt{\rho}}\frac{\partial^2
\sqrt{\rho}}{\partial x^2}
 = a \label{compensation1}\end{equation}
 In terms of $y (z) = \sqrt{\rho}$, $z = x - u_0 t$, then we have
 nonlinear equation
 \begin{equation}
 \frac{d^2 y}{dz^2} - \frac{a}{2} y + \frac{1}{2} y^3 = 0\label{nonlinear}
 \end{equation}
or multiplying with $y'$ and integrating once
\begin{equation}
 \left(\frac{d y}{dz}\right)^2 - \frac{a}{2} y^2 + \frac{1}{4} y^4 = 2b = const.\label{nonlinear1}
 \end{equation}
This equation has solution
\begin{equation}
 y = 2 p \, dn [p (x - u_0 t), \kappa]\label{dn}
 \end{equation}
 where $dn$ is the Jacobian elliptic function with the modulus
 $\kappa$, and $p$ is an arbitrary constant. It is connected with integration constants $a > 0$ and $b < 0$ by
 equations
\begin{equation}
 \frac{a}{2 p^2} = 1 + \kappa'^2,\,\,\,\,
\frac{b}{2 p^4} = - \kappa'^2
 \end{equation}
where $\kappa' = \sqrt{1 - \kappa^2}$ is the complimentary modulus
of the Jacobian elliptic function. These equations fix  relation
between the constants
\begin{equation}
a = 2 p^2\left(1 - \frac{b}{2 p^4}\right)\label{constants}
 \end{equation}
and the modulus $\kappa = \sqrt{1 + b/2 p^4}$. Then for  density
$\rho$ we have the travelling wave solution
\begin{equation}
 \rho (x,t) = 4 p^2 \, dn^2 [p(x - u_0 t), \kappa]\label{rho}\,.
 \end{equation}
With fixed potential (\ref{compensation1}) from equation
(\ref{madelung2}) and (\ref{velocity}) for  $u_0 = -
(2/\beta)\partial S/\partial x$, we have the Hamilton-Jacobi
equation
 \begin{equation}
-\frac{1}{\beta}\frac{\partial S}{\partial t} + \frac{1}{\beta^2}
\left(\frac{\partial S}{\partial x}\right)^2 + \frac{a}{2} =
0\label{Hamilton-Jacobi}
\end{equation}
which has solution in the form
 \begin{equation}
 S = \beta\left[S_0 - \frac{u_0}{2} x + \left(\frac{u_0^2}{4} + \frac{a}{2}\right)
 t \right]
\end{equation}
or using (\ref{constants})
\begin{equation}
 S = \beta\left[S_0 - \frac{u_0}{2} x + \left(\frac{u_0^2}{4} + p^2 (2 - \kappa^2)\right)
 t \right]
\label{S}
\end{equation}
For the degenerate case $\kappa = 1$ which corresponds to $b = 0$
and as follow $a = 2p^2$, the elliptic solution (\ref{rho}),
(\ref{S}) becomes the soliton
\begin{equation}
 \rho (x,t) = 4 p^2 \, sech^2 [p (x - u_0 t)]\label{rhosol}
 \end{equation}
\begin{equation}
 S = \beta\left[S_0 - \frac{u_0}{2} x + \left(\frac{u_0^2}{4} + p^2\right)
 t \right]
\label{Ssol}
\end{equation}

More general form of the travelling wave appears if we consider
solution of the system (\ref{cold1}),(\ref{cold2}) in the form
$
 u (x,t) = u(x - u_0 t)$, $\rho (x,t) = \rho(x - u_0 t)$.
 Then the first equation (\ref{cold1}) implies
$ \frac{d}{dz}[(u - u_0) \rho] = 0$, where $z = x - u_0 t$ and
\begin{equation}
u = u_0 + \frac{C}{\rho}\label{u}
\end{equation}
where $C = const.$ Substituting to the second equation
(\ref{cold2}) and integrating once we have
\begin{equation}
\frac{1}{2} \frac{C^2}{\rho^2} + \rho +
\left[\frac{1}{\rho}\,\frac{d^2 \rho}{d z^2} -
\frac{1}{2}\left(\frac{1}{\rho}\,\frac{d \rho}{d z}\right)^2
\right] = A = const.
\end{equation}
Multiplying with $\rho^2$ and differentiating once we have
(travelling wave form of the KdV equation)
\begin{equation}
\rho \left( \frac{d^3 \rho}{d z^3} + 3 \rho \frac{d \rho}{dz} - 2A
\frac{d \rho}{dz} \right) = 0
\end{equation}
which implies after one integration
\begin{equation}
 \frac{d^2 \rho}{d z^2} + \frac{3}{2} \rho^2 - 2A \rho  = B = const.\end{equation}
 This gives equation
\begin{equation}
 \left(\frac{d\rho}{d z}\right)^2 = - \rho^3 + 2A \rho^2 + 2B  \rho  + C^2\end{equation}
and solution
\begin{equation}
 \rho (x- u_0 t) = \alpha_1 + (\alpha_3 - \alpha_1) \,dn^2
 [\frac{1}{2}(\alpha_3 - \alpha_1)^{1/2} (x - u_0 t), \kappa]\label{periodic1}
 \end{equation}
 with modulus of elliptic function $\kappa^2 = (\alpha_3 - \alpha_2)/(\alpha_3 -
 \alpha_1)$, constants $\alpha_1, \,\alpha_2,\,\alpha_3$ and
\begin{equation}
 u (x- u_0 t) = u_0 - \frac{(\alpha_1
 \,\alpha_2\,\alpha_3)^{1/2}}{\rho}\label{periodic2}
 \end{equation}
This solution was first reported in \cite{GurevichKrylov88}. We
have particular reductions:

1)  $\alpha_1 = \alpha_2 = 0$ and corresponding $\kappa^2 = 1$ so
that (\ref{periodic1}) reduces to (\ref{rho}), with following
identification of parameters $\alpha_3 = 4 k^2$, $b = - \alpha_2
k^2/2$ and velocity $u = u_0$.

2) $\alpha_1 = \alpha_2 \neq 0$, then again $\kappa^2 = 1$ but
solution
\begin{equation}
 \rho (x- u_0 t) = \alpha_1 + (\alpha_3 - \alpha_1) \, sech^2
  [\frac{1}{2}(\alpha_3 - \alpha_1)^{1/2} (x - u_0 t)]\label{}
 \end{equation}
has nontrivial asymptotics. The physical value relevant for plasma
physics is $\alpha_1 = 1$ leading to $\lim_{|x|\rightarrow
\infty}\rho = 1$. Denoting $\alpha_3 \equiv \sigma^2$ we can write
it in the form
\begin{equation}
 \rho (x- u_0 t) = 1 + \frac{(\sigma^2 - 1)} {\cosh^2
  [\frac{\sqrt{\sigma^2 - 1}}{2} (x - u_0 t)]}\label{}
 \end{equation}

Using the above results we can now construct solutions of RNLS
(\ref{RNLSdefocusing}). Substituting  Eqs.(\ref{rho})(\ref{S}) to
Eq.(\ref{compfunc}) and changing parameter $k \equiv p/\beta$, we
have quasi-periodic solution
\begin{equation}
 \psi (x', t') = 2 \beta k \,\,dn[k (x' - u_0 t'), \kappa] e^{-i\{\phi_0 -
 \frac{u_0}{2}x' + [\frac{u_0^2}{4} + \beta^2 k^2 (2 - \kappa^2)]t'   \}}\label{periodicRNLS}
 \end{equation}
 where $x' = \beta x$, $t' = \beta t$.
 In the limit $\kappa = 1$ it gives the envelope soliton solution
\begin{equation}
 \psi (x', t') = 2 \beta k  \frac{e^{-i\{\phi_0 -
 \frac{u_0}{2}x' + [\frac{u_0^2}{4} + \beta^2 k^2 ]t'   \}}}{\cosh k (x' - u_0 t')}  \label{solitonRNLS}
 \end{equation}
For the reaction-diffusion system (\ref{RD}) correspondingly, we
have the "dissipative" periodic solution
\begin{equation}
 e^{(\pm)} (x', \tau) = \pm 2 \beta k \,\,dn[k (x' - v \tau), \kappa] e^{\pm\{\phi_0 -
 \frac{v}{2}x' + [\frac{v^2}{4} + k^2 (2 - \kappa^2)]\tau   \}}\label{periodicRD}
 \end{equation}
where velocity $v \equiv u_0/\beta$, $k \equiv p/\beta$ and the
dissipative analog of the envelope soliton
\begin{equation}
 e^{(\pm)} (x', \tau) = \pm 2 \beta k  \frac{e^{\pm \{\phi_0 -
 \frac{v}{2}x' + [\frac{v^2}{4} +  k^2 ]\tau   \}}}{\cosh k (x' - v \tau)}  \label{dissipatonRNLS}
 \end{equation}
the so called "dissipaton" solution \cite{PashaevLeeMPLA}

\section{Bilinear form and solitons. Trivial Boundary Conditions}

 The reaction-diffusion representation (\ref{RD}) of
 the RNLS (\ref{RNLSdefocusing}) is the key point for
 construction
 multisoliton solutions.
Due to algebraic similarity of RD system with NLS it is easy to
write the bilinear representation for it. Representing two real
functions $e^{(+)}$, $e^{(-)}$ in terms of three real functions
\begin{equation} e^{(\pm)} = 2 \beta \frac{G^{\pm}}{F} \label{rep}\end{equation}
we have the next bilinear system of equations
\begin{equation} (\pm D_{\tau} - D_{x}^2)(G^{(\pm)} \cdot F) = 0,\,\,
D_{x}^2(F \cdot F) = - 2 G^{(+)}G^{(-)}\label{bilinear}
\end{equation}
The corresponding solution of the RNLS (\ref{RNLSdefocusing}) has
\begin{equation} |\psi(x,t)|^2 = \rho = - e^{(+)}e^{(-)} =
2\beta^2 \frac{D_{x}^2(F \cdot F)}{F^2} = 4\beta^2
\frac{\partial^2 \ln F}{\partial x^2}\end{equation}

The one-dissipaton is given by the following solution of system
(\ref{bilinear}): $G^{\pm} = \pm e^{\eta^{\pm}_{1}}$, $F = 1 +
e^{\eta^{+}_{1} + \eta^{-}_{1} + \phi_{1,1}}$, $e^{\phi_{1,1}} =
(k^{+}_{1} + k^{-}_{1})^{-2}$, where $\eta^{\pm}_{1} \equiv
k^{\pm}_{1} x \pm (k^{\pm}_{1})^{2} \tau
  + \eta^{\pm (0)}_{1}$ and $k^{\pm}_{1}$, $\eta^{\pm (0)}_{1}$
are constants. In terms of redefined parameters, $k \equiv
(k^{+}_{1} + k^{-}_{1})/2$, $v \equiv -(k^{+}_{1} - k^{-}_{1})$ it
acquires the form (\ref{dissipatonRNLS}). In the space of
    parameters $(v, k)$ there exist  the
critical value  $v_{crit} =
 2k$ for  solution (\ref{1dissipaton}) so that when
$v < v_{crit}$,
   one has
$e^{\pm} \rightarrow 0$ at infinities, so the vanishing b.c. for
the dissipaton.
 At the critical value
 the solution is  a kink steady state  in
the moving frame
 $e^{\pm} = \pm k e^{\pm k \xi_{0}}
(1\, \mp \, \tanh \;k\xi )$, with constant asymptotics $e^{\pm}
\rightarrow \pm \, 2 k e^{\pm k \xi_{0}}$ for $x \rightarrow \mp
\infty$ and $e^{\pm} \rightarrow \pm \, 0$ for  $x \rightarrow \pm
\infty$.  In the  over-critical case $v > v_{crit}$, $e^{\pm}
\rightarrow \pm \infty$ for $x \rightarrow \mp \infty$ and
$e^{\pm} \rightarrow \pm \, 0$ for  $x \rightarrow \pm \infty$.
\par
For the two-dissipaton solution we have \begin{equation}
G^{\pm} =
\pm [e^{\eta^{\pm}_{1}} + e^{\eta^{\pm}_{2}} + ({\breve
k^{\pm\pm}_{12}\over k^{\pm\mp}_{21}k^{+-}_{11}})^{2}
e^{\eta^{+}_{1} + \eta^{-}_{1} + \eta^{\pm}_{2}} +
 ({\breve k^{\pm\pm}_{12}\over k^{\pm\mp}_{12}k^{+-}_{22}})^{2}
e^{\eta^{+}_{2} + \eta^{-}_{2} + \eta^{\pm}_{1}}] ;\end{equation}
\par
\begin{eqnarray} F = 1 + {e^{\eta^{+}_{1} + \eta^{-}_{1}} \over
(k^{+-}_{11})^{2}} + {e^{\eta^{+}_{1} + \eta^{-}_{2}} \over
(k^{+-}_{12})^{2}} + {e^{\eta^{+}_{2} + \eta^{-}_{1}} \over
(k^{+-}_{21})^{2}} + {e^{\eta^{+}_{2} + \eta^{-}_{2}} \over
(k^{+-}_{22})^{2}}\nonumber
\\
+ ({\breve k^{++}_{12}\breve k^{--}_{12} \over k^{+-}_{12}
k^{+-}_{21}k^{+-}_{11}k^{+-}_{22}})^{2} e^{\eta^{+}_{1} +
\eta^{-}_{1} + \eta^{+}_{2} + \eta^{-}_{2}},\end{eqnarray} where
$k^{ab}_{ij} \equiv k^{a}_{i} + k^{b}_{j},\,\,\,\,\,\,
 \breve k^{ab}_{ij} \equiv k^{a}_{i} - k^{b}_{j},$
$\eta^{\pm}_{i} \equiv k^{\pm}_{i} x \pm (k^{\pm}_{i})^{2} \tau
  + \eta^{\pm (0)}$. This solution shows the resonance character
  of dissipatons interaction \cite{PashaevLeeMPLA}

\section{Bilinear form and solitons. Non-trivial Boundary Conditions}

As was first noticed by Hirota, for NLS equation of defocusing
type with nonvanishing boundary conditions, the bilinear form of
equations should be modified \cite{Hirota}. After substituting
representation (\ref{rep}) to the system (\ref{RD}) the decoupling
system is choosen in the form
\begin{equation} (\pm D_{\tau} - D_{x}^2 + \lambda)(G^{\pm} \cdot F) = 0,\,\,
(D_{x}^2 - \lambda )(F \cdot F)  = - 2 G^{+}G^{-}\label{bilinear2}
\end{equation}
where we have introduced constant $\lambda$ to be determined.
Equation (\ref{rep}) and the second one of (\ref{bilinear2}) imply
$ - e^{(+)} e^{(-)} = -4\beta^2\left[ \frac{\lambda}{2} - (\ln
F)_{xx}\right]$. Expanding $G^{\pm}$ and $F$ in Hirota's power
series
\begin{equation}
G^{\pm} = \pm g^{\pm}_0 (1 + \epsilon g^{\pm}_1 + \epsilon^2
g^{\pm}_2 + ...),\,\,\,F = 1 + \epsilon f_1 + \epsilon^2 f_2 + ...
,
\end{equation}
and requiring $\lim_{|x| \rightarrow \infty} (\ln F)_{xx} = 0 $ we
have the boundary condition
\begin{equation}
\alpha_1 = \lim_{|x| \rightarrow \infty}[- e^{(+)} e^{(-)}] =
\lim_{|x| \rightarrow \infty} -4\beta^2\left[ \frac{\lambda}{2} -
(\ln F)_{xx}\right] = - 2\beta^2\lambda
\end{equation}
which fixes constant $\lambda = -\alpha_1/2\beta^2 $. In the zero
order approximation we have the system
\begin{equation} (\pm D_{\tau} - D_{x}^2 + \lambda)(g^{\pm}_0 \cdot 1) = 0,\,\,
(D_{x}^2 - \lambda )(1 \cdot 1)  =  2 g^{+}_0 g^{-}_0
\label{bilinear0}
\end{equation}
It has a solution in the form $g^{\pm}_0 =
\beta^{\pm}e^{\theta^{\pm}}$, where $\beta^2_0 \equiv \beta^{+}
\beta^{-} = - \lambda /2 =\alpha_1/4\beta^2$, $\theta^{\pm} = \pm
k x \pm (k^2 - \lambda) t$, $\beta^{\pm} = \beta_0 e^{\pm
\gamma_0}$ . Using following properties of Hirota's derivatives
\begin{equation} D_{x}(f g \cdot h) = \frac{\partial f}{\partial x} g h  +  f D_{x}(g \cdot h) \label{HirotaRule1}
\end{equation}
\begin{equation} D^2_{x}(f g \cdot h) = \frac{\partial^2 f}{\partial x^2} g h
+ 2 \frac{\partial f}{\partial x}D_{x}(g \cdot h) +  f D^2_{x}(g
\cdot h) \label{HirotaRule2}
\end{equation}
the bilinear system becomes
\begin{equation} (\mp D_{\tau} \pm 2k D_x + D_{x}^2)((1 + \epsilon g^{\pm}_1 +
\epsilon^2 g^{\pm}_2 + ...) \cdot (1 + \epsilon f_1 +
\epsilon^2 f_2 + ...)) = 0, \label{shiftbilinear1}
\end{equation}
\begin{equation} (D_{x}^2 + 2 \beta^2_0)((1 + \epsilon f_1 + ...) \cdot (1 + \epsilon f_1 +
...)) = 2 \beta^2_0 (1 + \epsilon g^{+}_1 +  ...)(1 + \epsilon
g^{-}_1 + ...). \label{shiftbilinear2}
\end{equation}

\subsection{One soliton solution}
In the first order we have the system
\begin{equation} (\mp \partial_{\tau} \pm 2k \partial_x + \partial_{x}^2)g^{\pm}_1 +
(\pm \partial_{\tau} \mp 2k \partial_x + \partial_{x}^2)f_1 = 0,
\label{order1-a}
\end{equation}
\begin{equation} (\partial_{x}^2 + 2 \beta^2_0) f_1  =
\beta^2_0 ( g^{+}_1 + g^{-}_1 ). \label{order1-b}
\end{equation}
Considering solution in the form $ g^{\pm}_1 = a^{\pm}_1
e^{\eta_1},\,\, f_1 = b_1 e^{\eta_1}$, where $\eta_1 = k_1 x +
\omega_1 \tau + \eta^{0}_1$ we get $ a^{+}_1 = \gamma_1 b_1$,
$a^{-} = \frac{1}{\gamma_1} b_1$, $\gamma_1 = (\omega_1 - 2 k k_1
+ k^2_1)/(\omega_1 - 2 k k_1 - k^2_1)$, where dispersion formula
is
\begin{equation}
\omega^{\pm}_1 = k_1 (2 k \pm \sqrt{k^2_1 + 4 \beta^2_0})
\end{equation}
It is worth to notice that in contrast to the dark soliton of the
defocusing NLS equation, in our case no restrictions on values of
$k_1$ appear.

Truncation of the Hirota expansion at this level gives one
dissipative soliton
\begin{equation}
e^{(\pm)} = \pm 2\beta\beta^{\pm} e^{\pm [k x + (k^2 + 2
\beta^2_0)\tau]} \frac{1 + \gamma^{\pm 1}_1 e^{\tilde\eta_1}}{1 +
 e^{\tilde\eta_1}}
\end{equation}
were we have absorbed constant $b_1$ to the exponential form
$\tilde\eta_1 = k_1 x + \omega_1 \tau + \eta^{0}_1 + \ln b_1$.
Then we have one soliton density
\begin{equation}
\rho  = - e^{(+)}e^{(-)} = \alpha_1 \frac{
 (1+\gamma_1e^{\tilde\eta_1})(1+\gamma^{-1}_1 e^{\tilde\eta_1})}{(1+e^{\tilde\eta_1})^2}
\end{equation}
 This solution can be represented in the form
\begin{equation}
e^{(\pm)} = \pm\sqrt{\alpha_1}\mu^{\pm 1}e^{\pm [k x + (k^2 + 2
\beta^2_0)\tau]}\left( \frac{\gamma^{\pm 1} +1}{2} +
\frac{\gamma^{\pm 1} -1}{2} \tanh \frac{\tilde\eta}{2} \right)
\end{equation}
or
\begin{equation}
e^{(+)} = +\sqrt{\alpha_1}\,\frac{\mu}{2}\, e^{+[k x + (k^2 + 2
\beta^2_0)\tau]}\left( {\gamma +1} + (\gamma -1) \tanh
\frac{\tilde\eta}{2} \right)
\end{equation}
\begin{equation}
e^{(-)} = -\sqrt{\alpha_1}\,\frac{1}{2\mu}\, e^{-[k x + (k^2 + 2
\beta^2_0)\tau]}\left( \frac{1}{\gamma} +1 + (\frac{1}{\gamma} -1)
\tanh \frac{\tilde\eta}{2} \right)
\end{equation}
and the product is
\begin{equation}
\rho  = - e^{(+)}e^{(-)} = \alpha_1\,\left[ 1 + \frac{(\gamma -
1)^2}{4 \gamma \cosh^2\frac{\tilde\eta}{2}} \right]
\end{equation}
with asymptotic $\lim_{|x|\rightarrow \infty} \rho \rightarrow
\alpha_1 $. Explicitly it is
\begin{equation}
\rho  = - e^{(+)}e^{(-)} = \alpha_1\,\left[ 1 + \frac{k^2_1}{4
\beta^2_0} \, sech^2\frac{k_1}{2}\left(x + (2 k \pm \sqrt{k_1^2 +
4\beta^2_0})\tau + x_0\right) \right]
\end{equation}
where $\beta^2_0 = 1/(4\beta^2 \alpha_1)$. For the velocity field
we have
\begin{equation}
u = \frac{e^{(-)}_x}{e^{(-)}}-\frac{e^{(+)}_x}{e^{(+)}} = - 2k -
\frac{(\gamma^2 - 1)k_1}{(\gamma -1)^2 + 4 \gamma \cosh^2
\frac{\tilde\eta}{2}}\label{1velocity}
\end{equation}

Let us consider particular solution for $k = 0$, then as follows
\begin{equation}\omega_1 = \pm k_1 \sqrt{k^2_1 + 4 \beta^2_0}\end{equation} is the
Bogolubov dispersion from the theory of superfluidity of a weakly
non-ideal Bose gas. For $k_1 >> 1$ it is of the non-relativistic
free particles form $\omega_1 \approx k^2_1$, while for $k_1 << 1$
it is of the relativistic collective  form $\omega_1 \approx 2
\beta_0 k_1$. Solution for the + sign of the dispersion has the
form
\begin{equation}
e^{(+)} = +\frac{\sqrt{\alpha_1} \mu}{v - \sqrt{v^2 -
4\beta^2_0}}\, e^{+ 2 \beta^2_0 \tau}\left( v  + \sqrt{v^2 -
4\beta^2_0} \tanh \frac{\sqrt{v^2 - 4\beta^2_0}}{2}(x + v \tau +
x_0) \right)
\end{equation}
\begin{equation}
e^{(-)} = -\frac{\sqrt{\alpha_1}\,/\mu }{v + \sqrt{v^2 -
4\beta^2_0}}\, e^{- 2 \beta^2_0 \tau}\left( v  - \sqrt{v^2 -
4\beta^2_0} \tanh \frac{\sqrt{v^2 - 4\beta^2_0}}{2}(x + v \tau +
x_0) \right)
\end{equation}
and the density is
\begin{equation}
\rho  = - e^{(+)}e^{(-)} = \alpha_1\,\left[ 1 + \frac{v^2 - 4
\beta^2_0}{4 \beta^2_0} \, sech^2  \frac{\sqrt{v^2 -
4\beta^2_0}}{2}(x + v \tau + x_0)\right]
\end{equation}
It shows that velocity of soliton is bounded from below by modulus
$|v| > 2 \beta_0$, so the soliton has the "tachionic" character.
These results show first that in contrast with defocusing NLS with
soliton's  velocity bounded from above (subsonic type), the RNLS
soliton has velocity bounded from below (supersonic type). Another
difference is that soliton of defocusing (repulsive) NLS is the
hole (bubble) like excitation with $|\psi|^2 = \rho < 1$, while
for the RNLS soliton we have the wall like form $\rho >1$.

\subsection{Two soliton solution}
To construct two soliton solution following Hirota \cite{Hirota}
we consider
\begin{equation}
g^{(\pm)}_1 = a^{\pm}_1 e^{\eta_1} + a^{\pm}_2
e^{\eta_2},\,\,\,f_1 = e^{\eta_1} + e^{\eta_2}
\end{equation}
Substituting to bilinear equations
\begin{eqnarray}
(\mp D_{\tau} \pm 2k D_x + D_{x}^2)(a^{\pm}_1 e^{\eta_1} +
a^{\pm}_2 e^{\eta_2}) \cdot 1 + 1\cdot (e^{\eta_1} + e^{\eta_2}) =
0,
\\ 2(D_{x}^2 + 2 \beta^2_0)(1 \cdot (e^{\eta_1} + e^{\eta_2})) =
2 \beta^2_0 (a^{+}_1 e^{\eta_1} + a^{+}_2 e^{\eta_2}+ a^{-}_1
e^{\eta_1} + a^{-}_2 e^{\eta_2})
\end{eqnarray}
we have the system
\begin{eqnarray}
a^{\pm}_1(\mp \partial_{\tau} \pm 2k \partial_x +
\partial^2_{x}) e^{\eta_1} +  (\pm \partial_{\tau} \mp 2k \partial_x +
\partial^2_{x})e^{\eta_1}  + \nonumber\\
a^{\pm}_2(\mp \partial_{\tau} \pm 2k \partial_x +
\partial^2_{x}) e^{\eta_2} +  (\pm \partial_{\tau} \mp 2k \partial_x +
\partial^2_{x})e^{\eta_2}
 = 0,
\\ (k_1^2 + 2 \beta^2_0)(e^{\eta_1} + e^{\eta_2}) =
 \beta^2_0 [(a^{+}_1 + a^{-}_1)
e^{\eta_1} + (a^{+}_2 + a^{-}_2)e^{\eta_2}]
\end{eqnarray}
Using dispersion
\begin{equation}
\omega^{\pm}_i = k_i \left(2 k \pm \sqrt{k^2_i + 4
\beta^2_0}\right),\,\,\,(i=1,2) \label{2dispersion}\end{equation}
we have
\begin{equation}
a^+_i = \frac{(\omega_i - 2 k k_i)+ k^2_i}{(\omega_i - 2 k k_i)-
k^2_i} \equiv e^{+\phi_i},\,\,\,a^-_i = \frac{(\omega_i - 2 k
k_i)- k^2_i}{(\omega_i - 2 k k_i)+ k^2_i} = \frac{1}{a^+_i} =
e^{-\phi_i}
\end{equation}
The last relations imply
\begin{equation}
(\omega_i - 2 k k_i) = k^2_i \coth \frac{\phi_i}{2}
\end{equation}
and
\begin{equation}
k_i = 2 \beta_0 \sinh \frac{\phi_i}{2}
\end{equation}
so that
\begin{equation}
(\omega_i - 2 k k_i) = 2 \beta^2_0 \sinh \phi_i
\end{equation}
We note that two signs in dispersion (\ref{2dispersion})
correspond in the above formulas to the simple replacement $\phi_i
\rightarrow - \phi_i$. First we restrict consideration with the
same sign for both frequences. In the next order we have the
system
\begin{eqnarray}
(\mp D_{\tau} \pm 2k D_x + D_{x}^2)(g^{\pm}_2\cdot 1 +
g^{\pm}_1\cdot f_1 + 1\cdot f_2) = 0,
\\ (D_{x}^2 + 2 \beta^2_0)(2 \cdot f_2 + f_1\cdot f_1) =
2 \beta^2_0 (g^+_2 + g^-_2 + g^+_1 g^-_1)
\end{eqnarray}
or for the first equation
\begin{eqnarray}
(\mp \partial_{\tau} \pm 2k \partial_x +
\partial_{x}^2)g^{\pm}_2 + (\pm \partial_{\tau} \mp 2k \partial_x +
\partial_{x}^2)f_2 \nonumber \\+ \{ a^{\pm}_1 [\mp (\omega_1 - \omega_2) \pm 2k (k_1-k_2) +
(k_1-k_2)^2] \nonumber\\
+ a^{\pm}_2 [\mp (\omega_2 - \omega_1) \pm 2k (k_2-k_1) +
(k_1-k_2)^2] \} e^{\eta_1 + \eta_2} = 0,
\end{eqnarray}
It implies solution in the form
\begin{equation}
g^{\pm}_2 = a^{\pm}_{12} e^{\eta_1 + \eta_2},\,\,\,f_2 = b_{12}
e^{\eta_1 + \eta_2}
\end{equation}
Then the second equation of the system implies relations
\begin{equation}
a^{+}_{12} = a^+_1 a^+_2 b_{12} = e^{(\phi_1 + \phi_2)}
b_{12},\,\,\,a^{-}_{12} = a^-_1 a^-_2 b_{12} = e^{-(\phi_1 +
\phi_2)} b_{12}
\end{equation}
so that
\begin{equation}
b_{12} =
\frac{\sinh^2\frac{\phi_1-\phi_2}{4}}{\sinh^2\frac{\phi_1+\phi_2}{4}}
\end{equation}
and the first equation is satisfied automatically. As a result we
have solution
\begin{equation}
e^{(\pm)} = \pm 2\beta\frac{g^{\pm}_0 (1 + g^{\pm}_1 +
g^{\pm}_2)}{1 + f_1 + f_2}
\end{equation}
or
\begin{equation}
e^{(\pm)} = \pm 2\beta\frac{g^{\pm}_0 (1 + e^{\eta_1 \pm \phi_1} +
e^{\eta_2 \pm \phi_2} + b_{12}e^{\eta_1 + \eta_2 \pm (\phi_1 +
\phi_2)}}{1 + e^{\eta_1} +e^{\eta_2} + b_{12}e^{\eta_1 + \eta_2}}
\end{equation}
For particular parametrization $\beta_0 = 1/2$, implying $\alpha_1
= 1$, we have two soliton solution for density $\rho$,
\begin{equation}
\rho = \frac{A_+ A_-}{[\sinh^2 \frac{\phi_1 + \phi_2}{4}(1 +
e^{\eta_1} + e^{\eta_2}) + \sinh^2 \frac{\phi_1 - \phi_2}{4}
e^{\eta_1 + \eta_2}]^2}
\end{equation}
where
\begin{equation}
A_{\pm} = \sinh^2 \frac{\phi_1 + \phi_2}{4}(1 + e^{\eta_1 \pm
\phi_1} + e^{\eta_2 \pm \phi_2}) + \sinh^2 \frac{\phi_1 -
\phi_2}{4} e^{\eta_1 + \eta_2 \pm (\phi_1 + \phi_2)},
\end{equation}
\begin{equation}\eta_i = \sinh \frac{\phi_i}{2}x + [2 k \sinh \frac{\phi_i}{2}
+ \frac{1}{2}\sinh \phi_i]\tau + \eta^{(0)}_i, \,\,\,(i = 1,2)
\end{equation}
If one of the parameters $\phi_i$ is vanishing or if $\phi_1 =
\phi_2$, the solution is reduced to the one soliton form. For
example if $\phi_2 = 0$ then
\begin{equation}
\rho = 1 + \frac{\sinh^2 \frac{\phi_1}{2}}{\cosh^2
\frac{\eta_1}{2}}\label{1soliton}
\end{equation}
Analyzing two soliton solution in the soliton's moving frames we
can see that it describes collision of two solitons type
(\ref{1soliton}), moving in the same direction with the initial
position shifts
\begin{equation}
\Delta x_i = (-1)^{i-1}\frac{2}{\sinh \frac{\phi_i}{2}}\ln
\frac{\sinh \frac{\phi_1 - \phi_2}{4}}{\sinh \frac{\phi_1 +
\phi_2}{4}},\,\,\,\,(i=1,2) \label{position}
\end{equation}
so that $\sinh \frac{\phi_1}{2}\Delta x_1 + \sinh \frac{\phi_2}{2}
\Delta x_2 = 0$.

The different form of two soliton solution we obtain if we choose
opposite signs for frequences (\ref{2dispersion}), so that
\begin{eqnarray}
\omega^{+}_1 &=& 2\beta_0 (2 k \sinh \frac{\phi_1}{2} + \beta_0
\sinh \phi_1),\\\omega^{-}_2 &=& 2\beta_0 (2 k \sinh
\frac{\phi_1}{2} - \beta_0 \sinh \phi_1)
\end{eqnarray}
Then $a^{\pm}_1 = e^{\pm \phi_1}$, $a^{\pm}_1 = e^{\mp \phi_1}$,
\begin{eqnarray}
 a^+_{12} = a^+_1 a^+_2 b_{12} = e^{\phi_1 -\phi_2} b_{12},\\
a^-_{12} = a^-_1 a^-_2 b_{12} = e^{-\phi_1 +\phi_2} b_{12}
\end{eqnarray}
and
\begin{equation}
b_{12} = \frac{\cosh^2 \frac{\phi_1 + \phi_2}{4}}{\cosh^2
\frac{\phi_1 - \phi_2}{4}}
\end{equation}
For $\beta_0 = 1/2$ two soliton solution is
\begin{equation}
\rho = \frac{B_+ B_-}{[\cosh^2 \frac{\phi_1 - \phi_2}{4}(1 +
e^{\eta_1} + e^{\eta_2}) + \cosh^2 \frac{\phi_1 + \phi_2}{4}
e^{\eta_1 + \eta_2}]^2}
\end{equation}
where
\begin{equation}
B_{\pm} = \cosh^2 \frac{\phi_1 - \phi_2}{4}(1 + e^{\eta_1 \pm
\phi_1} + e^{\eta_2 \mp \phi_2}) + \cosh^2 \frac{\phi_1 +
\phi_2}{4} e^{\eta_1 + \eta_2 \pm \phi_1 \mp \phi_2},
\end{equation}
and
\begin{eqnarray}
 \eta_{1} = \sinh \frac{\phi_1}{2} (x - x_1) + [2 k \sinh \frac{\phi_1}{2} + \frac{1}{2}\sinh \phi_1] \tau,\\
\eta_{2} = \sinh \frac{\phi_2}{2} (x - x_2) + [2 k \sinh
\frac{\phi_2}{2} - \frac{1}{2}\sinh \phi_2 ] \tau
\end{eqnarray}
It describes collision of two solitons in the form
(\ref{1soliton}), moving in opposite direction with initial
position shifts
\begin{equation}
\Delta x_i = (-1)^{i-1}\frac{2}{\sinh \frac{\phi_i}{2}}\ln
\frac{\cosh \frac{\phi_1 + \phi_2}{4}}{\cosh \frac{\phi_1 -
\phi_2}{4}},\,\,\,\,(i=1,2) \label{position}
\end{equation}
In Fig.1 we show 3D plot of this solution.
\begin{figure}[h]
\begin{center}
\epsfig{figure=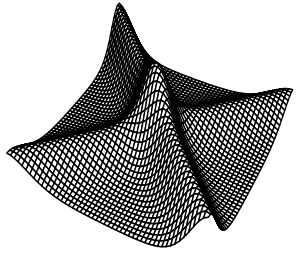,height=4cm,width=4cm}
\end{center}
\caption{Two Soliton Scattering} \label{Bound4}
\end{figure}

For the velocity field we have
\begin{eqnarray}
u = \frac{e^{(-)}_x}{e^{(-)}} - \frac{e^{(+)}_x}{e^{(+)}}= - 2k +
\hskip6cm\\
+\frac{\cosh^2 \frac{\phi_1 - \phi_2}{4}\, a_- + \cosh^2
\frac{\phi_1 - \phi_2}{4}\, b_-}{\cosh^2 \frac{\phi_1 - \phi_2}{4}
(1+ e^{\eta_1-\phi_1} + e^{\eta_2+\phi_2}) + \cosh^2 \frac{\phi_1
+ \phi_2}{4} e^{\eta_1 + \eta_2 -\phi_1 + \phi_2}}\nonumber\\-
\frac{\cosh^2 \frac{\phi_1 - \phi_2}{4}\, a_+ + \cosh^2
\frac{\phi_1 - \phi_2}{4}\, b_+}{\cosh^2 \frac{\phi_1 - \phi_2}{4}
(1+ e^{\eta_1+\phi_1} + e^{\eta_2-\phi_2}) + \cosh^2 \frac{\phi_1
+ \phi_2}{4} e^{\eta_1 + \eta_2 +\phi_1 - \phi_2}}\nonumber
\end{eqnarray}
where
\begin{equation}
a_\mp = (\sinh \frac{\phi_1}{2}e^{\eta_1 \mp \phi_1} + \sinh
\frac{\phi_2}{2}e^{\eta_2 \pm \phi_2})
\end{equation}
\begin{equation}
b_\mp = (\sinh \frac{\phi_1}{2} + \sinh \frac{\phi_2}{2})e^{\eta_1
+ \eta_2 \mp \phi_1 \pm \phi_2}
\end{equation}

It has the same phase shift and describe collision of two solitons
in the form
\begin{equation}
u = -2k - \frac{k_i \sinh \phi_i}{2 \cosh \frac{\eta_i +
\phi_i}{2} \cosh \frac{\eta_i - \phi_i}{2}}\,,\, (i =1,2).
\end{equation}

\section{Acknowledgments} This work was supported in part by the
Institute of Mathematics, Academia Sinica, Taipei, Taiwan, and the
Izmir Institute of Technology, Izmir, Turkey, BAP project 2005
IYTE 08.

\end{document}